\documentclass[twocolumn,superscriptaddress,aps,prl]{revtex4-1}
\usepackage{amssymb,amsmath}
\usepackage{graphicx}
\usepackage[colorlinks,linkcolor=blue,citecolor=blue,urlcolor=blue]{hyperref}

\begin{document}

\title{Spin-selective metal insulator transition in two dimensions}

\author{Shashikant Singh Kunwar}
\affiliation{Department of Physics, Indian Institute of Technology, Madras,
Chennai 600036, India}
\author{Prabuddha B. Chakraborty}
\affiliation{Indian Statistical Institute, Chennai Centre, 110 N. Manickam Road, Rajaram Mehta Nagar, Aminjikarai, Chennai 600029, India} 
\author{R. Narayanan}
\affiliation{Department of Physics, Indian Institute of Technology, Madras,
Chennai 600036, India}

\date{\today}

\begin{abstract}
The phenomenon of Anderson localization wherein non-interacting electrons are localized by quenched impurities is a subject matter that has been extremely well studied. However, localization transition under the combined influence of interaction and quenched disorder is less well understood. In this context we study the localization transition in a two-dimensional Hubbard model under the influence of a spin-selective disorder i.e, disorder which is operational on just one of the spin-species. The model is analyzed by laying recourse to a Quantum Monte Carlo based scheme. Using this approach we show the possibility of a metal-insulator transition at densities away from half-filling.

\end{abstract}

\maketitle

\paragraph{Introduction:}
The paradigmatic scaling theory\cite{gangoffour} of localization\cite{anderson} attests to the absence of a metallic state for a system of non-interacting 
disordered electrons belonging to certain symmetry classes \cite{altland,evers_mirlin} in $d\le2$ dimensions.  However, 
theoretically the existence of a metallic state in low-dimensional ($d\le 2$) electronic system under the combined effects of interaction and disorder
remains an open problem. The development of such a theory is particularly important as intriguing experiments on two dimensional MOSFET devices with extremely high mobility seem to indicate the existence of a metallic state and thus a possible Metal to Insulator Transition (MIT) as a function of interaction  strength \cite{kravchenko, kravchenko2004}. Furthermore, in the recent past, the study of phases and phase transitions in interacting disordered systems in low-dimensions have garnered further impetus with the advent of the field of Many Body Localization (MBL) wherein an interplay of disorder and short-range interactions leads to a perfectly insulating state at finite temperatures \cite{basko, gornyi, oganesyan}. 

In this Letter we concern ourselves with the question of whether the combined effects of disorder and short-range interactions could trigger a metallic state in a 
two dimensional fermionic system. To do so we focus on a repulsive Hubbard model in the presence of spin-dependent disorder. Much like the MBL states which can be 
probed by laying recourse to cold atomic systems \cite{luschen,guo}, spin-dependent disorder can also be precision engineered in similar systems by using laser beams of two different polarizations to create a speckle type of disorder which acts differently on particles with different spin-polarizations \cite{prabudhaSSThermo}.   
 
  The starting point of our analysis is the disordered Hubbard model defined on the square lattice: 
 \begin{equation}
\begin{aligned}
H= - \sum_{\langle i,j \rangle,\sigma} t ({c_{i,\sigma}^{\dagger}}c_{j,\sigma}+h.c)+\sum_{i,\sigma}\mu_{i,\sigma}n_{i,\sigma} \\ + U \sum_{i}\left (n_{i,\uparrow}-\frac{1}{2}\right )\left(n_{i,\downarrow}-\frac{1}{2}\right).
\label{1}
\end{aligned} 
\end{equation}
 Here, in Eq.~\ref{1}, $t$ is the nearest neighbour hopping amplitude which has been set to unity. 
 The on-site interaction is repulsive, i.e., $U>0$.  Furthermore, $c_{i,\sigma}^{\dagger}$, and $c_{i,\sigma}$ are the fermionic creation and 
 annihilation operators defined on site $i$ and carry the spin index $\sigma$. The effect of {\em spin-selective disorder} is incorporated by making the 
 random chemical potential {\em spin-dependent}. In our realization, in the limit $U=0$, the $\sigma =\downarrow$ fermions feels the effect of a random 
 chemical potential whereas $\sigma =\uparrow$ fermionic species moves in the background of a non-random chemical potential. More precisely, the 
 chemical potential seen by the $\sigma=\downarrow$ species of fermions $\mu_{i,\downarrow}$ is uniformly sampled from the range $\left[ -\frac{\Delta_\downarrow}{2}, \frac{\Delta_\downarrow}{2}\right]$.  This is at odds with a conventionally disordered system wherein both the spin-species sees the same disordered chemical potential. 
 
 The question of whether the interplay of interactions and conventional disorder can lead to a MIT has been tackled by a variety of means. Analytically, the approach has 
 centred on the field-theoretical adaptation \cite{fnklstein, dbtrk1,dbtrk2} of Wegner's sigma model for non-interacting electrons \cite{wegner}. However, this approach leads to Renormalization Group (RG) trajectories that flow to large coupling. To make contact with experiments on MOSFET devices \cite{kravchenko,kravchenko2004}, in a significant later development \cite{fnklsteinpunnose}, the field theory \cite{fnklstein, dbtrk1,dbtrk2} was adapted to account for valley degeneracy in semiconductors. Indeed it was shown that
 a interaction driven MIT indeed obtains in the limit  $n_{\rm v} \rightarrow \infty$, (where $n_{\rm v}$ is the number of valleys in a degenerate semiconductor) \cite{fnklsteinpunnose}. However, the $n_{\rm v} \rightarrow \infty$ limit brings forth  the question of applicability of this theory to the case of the Si- MOSFET heterostructures where it is well known that the number of valleys is two.

 Another oft-used track to theoretically study the question of metal-insulator transition in a two dimensional interacting disordered electron system is by laying 
 recourse to Quantum Monte-Carlo based simulation methods \cite{QMC1981,SANTOS2003}. By using a Determinant Quantum Monte-Carlo Scheme (DQMC) Denteneer et al. 
 \cite{denteneer99,denteneer2001} has shown the existence of a MIT in a Hubbard model at quarter filling. However, progress using this technique is stymied due to the so-called 
 sign-problem\cite{Troyer2005}. 
 
  The effect of spin-dependent disorder on a repulsive Hubbard model was investigated within a DMFT framework \cite{DMFTSS}.  
  Very intriguingly, these investigations pointed to the existence of a novel spin-selective localized phase wherein one of the spin-species gets localized due to the disorder potential whereas the other spin species retain its itinerant character. Apart from the spin-selective localized phase the model also hosts the putative correlated metallic phase at small values of both interactions and disorder and a disordered Mott insulator at large values of the interaction. Furthermore, it was shown in Ref.~\cite{DMFTSS} that the transition from the correlated metal to the spin-selective localized phase follows the conventional Anderson localization scenario, whereas the transition from the spin-selective localized phase to the Mott Insulator is of the Falicov-Kimball type \cite{Falicov}.
  
In this paper, by using DQMC based simulations we report the possibility of a {\em novel interaction driven  de-localization} transition in a Hubbard-model 
 in the presence of a spin-selective disorder (see Eq.~\ref{1}). We show that this delocalization transition exists for filling fractions that are away from half-filling and 
 happens at very large values of interaction strength. We further 
 show that this de-localization transition resembles the Finkelstein scenario wherein the ferromagnetic spin-susceptibility gets enhanced close to the MIT.

\paragraph{Observables:}
The temperature dependent dc conductivity is the primary observable in the search of MIT in the model given by Eq.~\ref{1}. Since the DQMC procedure we follow is 
very standard and straightforward we present the discussion of the same in the Supplementary material \cite{Supplemental}.
The dc conductivity $\sigma_{{\rm dc}}$ is evaluated by means of a standard relation that connects it to a current-current correlation function $\Lambda(q,\tau)$: \cite{trivedi_scalettar}:
\begin{equation}
\sigma_{{\rm dc}} = \frac{\beta^2}{\pi}\Lambda(q=0,\tau=\beta/2). 
\label{eq:2}
\end{equation}
Here, in Eq.~\ref{eq:2}, $\Lambda(q,\tau) = \langle j_{x}(q,\tau)j_{x}(-q,\tau) \rangle$ and $\beta$ is the inverse temperature. Furthermore, $j_x(q,\tau)$ is the current operator given 
by $$j_{x}(r)=\frac{i e a t}{h}\sum_{\sigma} (c^{\dagger}_{r+e_{x}, \sigma}  c_{r,\sigma} - c^{\dagger}_{r,\sigma}  c_{r+e_{x}, \sigma}).$$ 
Following \cite{prabuddha_mag} we also evaluate the $\chi_{zz} (0,0) $ or the ferromagnetic component of the susceptibility. The susceptibility is given by the relation $\chi_{zz}(q)= \beta S(q)$, with the spin structure factor $S(q)$ given by: 
\begin{equation}
S(q) = \sum_r e^{iq.r} \langle S^z(R_i+r) S^z(R_i)\rangle.
\label{eq:susc}
\end{equation}
where $S^z (R_i) = n_{i\uparrow}-n_{i\downarrow}$.    
In the rest of the paper we follow a system of natural units.

\paragraph{Results:}

\begin{figure}[!htbp]
\centering
\includegraphics[width=0.5\textwidth,angle=0]{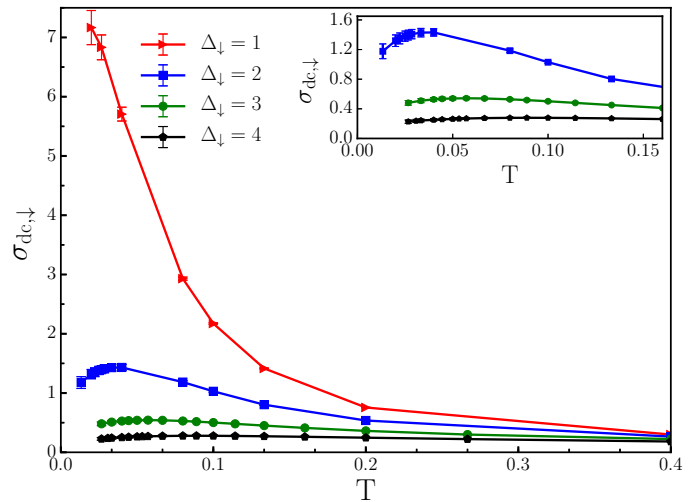}
\caption{The temperature dependence of the DC conductivity for spin down fermions as a function of disorder $\Delta_{\downarrow}$ for $U = 0$.}
\label{U0}
\end{figure}

{\it \underline{The case of half-filling:}}
 As a starting point of our analysis we concentrate on the situation 
of the half-filled lattice. The condition of half-filling is achieved by putting the chemical potential $\mu=0$ in Eq.~\ref{1}. Furthermore, 
as a test case we first put the repulsive interaction $U=0$. As expected the electrons with $\sigma = \downarrow$  that move in a disordered landscape 
gets localized by the presence of the spin-dependent impurities. This is clearly seen in Fig.~\ref{U0}, where as a  function of temperature ${\rm T}$ the 
$\sigma_{\rm dc}$ of the down spins show an insulating behavior, indicative of localization due to disorder. The onset temperature where one witnesses a down-turn in conductivity increases with increasing disorder strength, in accordance with Ref.~\cite{anderson}. 

\begin{figure}[!htbp]
\includegraphics[width=0.52\textwidth,angle=0]{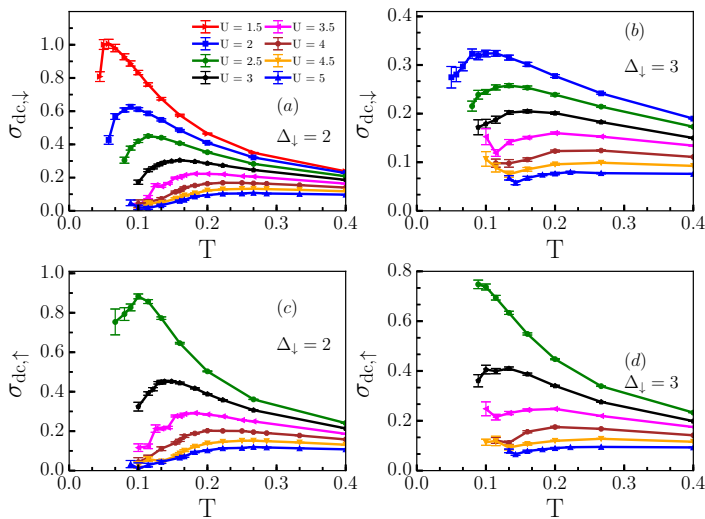}
\caption{The temperature dependence of DC conductivity for spin down particles $\sigma_{dc}^{\downarrow}$ (Fig.~\ref{halfspindn}(a), and Fig.~\ref{halfspindn}(b)) and spin up (Fig.~\ref{halfspindn}(c), and Fig.~\ref{halfspindn}(d)) as a function of interaction strength $U$ at disorder strength $\Delta_{\downarrow} = 2$ and $\Delta_{\downarrow} = 3$.}
\label{halfspindn}
\end{figure}

Now, turning on a non-zero interaction $U$, we see that the onset of the insulating phase in the spin-down sector happens at 
a higher value of temperature as compared to the non-interacting $U=0$ case, (see Fig.~\ref{halfspindn}(a), and Fig.~\ref{halfspindn}(b)).
 In similar vein, due to the presence of a non-zero interaction strength the spin-up particles will also see the effect of a "{\bf renormalized}" disorder strength. Thus, the spin-up particles also localize through an Anderson type mechanism which is clearly depicted in Fig.~\ref{halfspindn}(c) and Fig.~\ref{halfspindn}(d). 

Also, as seen from Fig.~\ref{halfspindn}, it is very clear that increasing either $U$ or the disorder strength $\Delta_\downarrow$ clearly enhances the insulating behavior. Furthermore, there seems to be no evidence of an interaction driven metal-insulator transition. Our results for the half-filled case should 
be contrasted to the DMFT results on the Hubbard model in the presence of spin-selective disorder \cite{DMFTSS}: For small values of interaction, Ref.~\cite{DMFTSS} report the existence of a spin-selective localized phase wherein one species of fermions gets localized with increasing disorder whereas the other species remain itinerant. In contrast in our simulations the addition of a small interaction always localizes both the species of spins. This variance in the behavior between the DMFT and the QMC results may be attributed to the special role played by dimensionality: Our QMC simulations are in $d=2$ whereas the DMFT results increasingly gets better for larger dimensions. 

%

\begin{figure}[!htbp]
\centering
\includegraphics[width=0.5\textwidth,angle=0]{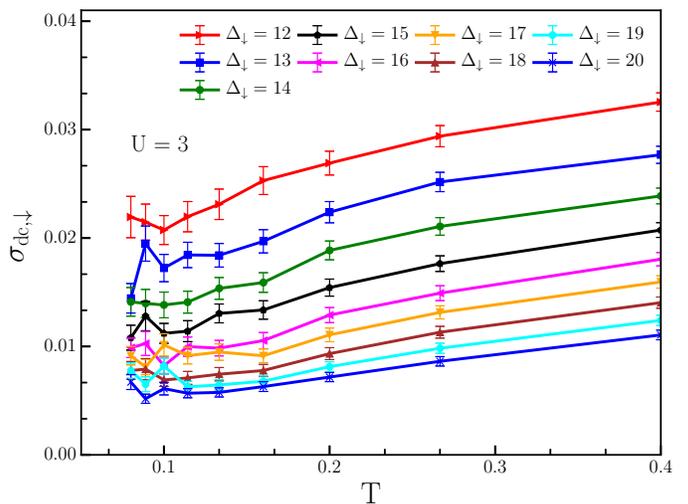}
\caption{DC conductivity for spin down particles $\sigma_{dc}^{\downarrow}$ particles as a function of disorder $\Delta_{\downarrow}$ at interaction $U = 3$.}
\label{quarterspindn}
\end{figure}

\begin{figure}[!htbp]
 \centering
 \includegraphics[width=0.45\textwidth]{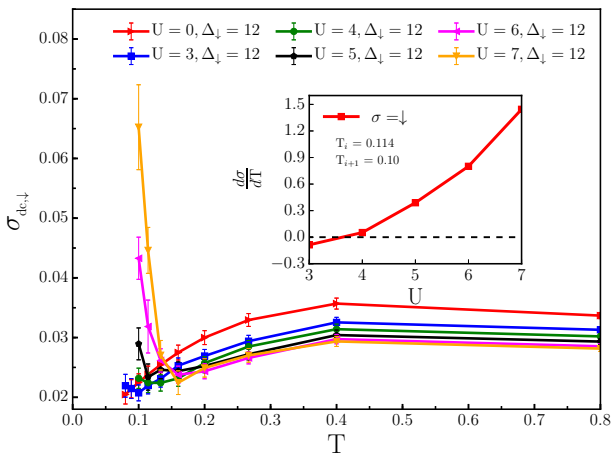}
 \caption{DC conductivity for spin down particles as a function of temperature at fixed disorder $\Delta_{\downarrow} = 12$ at various values of interaction  $U$ . Increasing interaction strength tends to delocalize the spin down particles. The inset depicts the sign-change of the slope $\frac{d\sigma}{dT}$ as a function of $U$ for the spin-down particles.}
 \label{quarterspindninter}
 \end{figure}

 {\it \underline{The case away from half-filling:}}
   Now, we turn our attention  to densities away from half-filling. As a prototypical example, in Fig.~\ref{quarterspindn}-to-Fig.~\ref{quarterspindninter} we illustrate the behavior of the conductivity as a function of temperature for the case of the quarter-filled lattice: More specifically, Fig.~\ref{quarterspindn} illustrates the behavior of the conductivity for the down-spin band as a function of the disorder strength. Just like in the half-filled case discussed above the conductivity decreases with decreasing temperature for all values of disorder strength $\Delta_{\downarrow}$ whilst keeping the interaction fixed.

\begin{figure}[!htbp]
\includegraphics[width=0.5\textwidth]{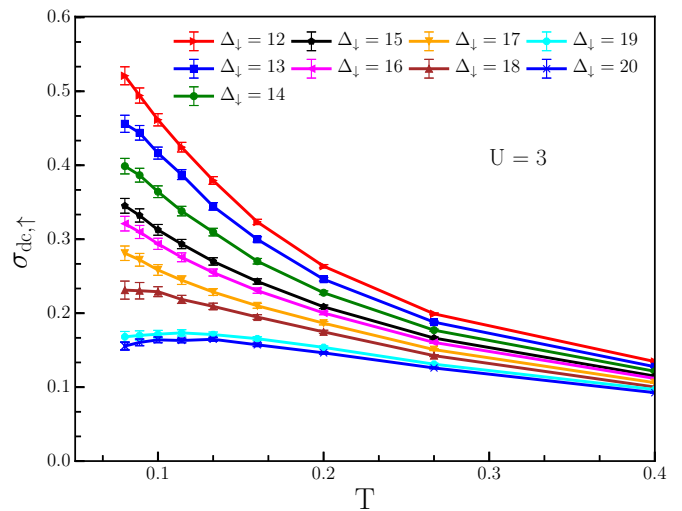}
\caption{Variation of DC conductivity of spin up particles as a function of disorder $\Delta_{\downarrow}$ at the fixed interaction $U = 3$. The behaviour of dc conductivity signals a disorder-driven metal-insulator transition}
\label{quarterspinup}
\end{figure}

Furthermore, Fig.~\ref{quarterspindn} clearly illustrates that the insulating behavior gets more pronounced with increasing disorder strength. In contrast, as seen in 
Fig~\ref{quarterspinup}, the spin-up particles that do not see the disordered environment at the bare-level, remain metallic for lower values of disorder and lower values of interaction. More specifically, the system shows metallic behavior, (increasing conductivity with decreasing temperature) for low enough 
values of disorder strength, (at $U=3$ for all values of $\Delta_{\downarrow} < 18$). However, as clearly seen in Fig.~\ref{quarterspinup}, for larger values of the disorder strength we observe evidence of an insulating behavior (decreasing conductivity with decreasing temperature). This clearly points to the existence of a disorder driven metal-insulator transition for the up-spin band at a critical value of disorder $\Delta^c_{\downarrow}$. 

\begin{figure}[!htbp]
 \centering
 \includegraphics[width=0.45\textwidth]{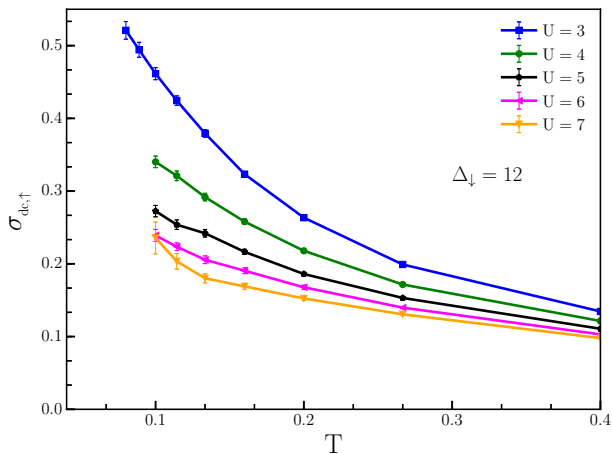}
 \caption{Temperature variation of DC conductivity for spin up particles at $\Delta_{\downarrow} = 12$ as a function of interaction strength. The sole effects of an interaction is to make the spin-up particles less metallic.}
 \label{quarterspinupinter}
 \end{figure}

We now turn our attention to the fate of the spin-up electrons as a function of increasing interaction strength. An example is seen in
Fig.~\ref{quarterspinupinter}, wherein for $\Delta_{\downarrow} =12$ we have plotted the behavior of $\sigma_{\rm{dc},\uparrow}$ as a function of temperature $T$ for various interaction strength $U$. As long as $\Delta_{\downarrow} < \Delta^c_{\downarrow}$, an increase in $U$ just leads to a worse metal. This implies that increasing $U$ just increases the effective disorder experienced by the minority band (spin-up particles). However, more importantly, upto the interaction strengths we can go into, before running into sign issues \cite{Troyer2005,SANTOS2003}, we see no evidence of an interaction driven metal-insulator transition for the up spin species. 
 
Now we turn our attention to the most striking result as evidenced in Fig.~\ref{quarterspindninter}. As can be clearly seen at low values of interaction 
strength at a fixed value of disorder strength ($\Delta_\downarrow=12$), the conductivity $\sigma_{\rm{dc},\downarrow}$ conforms to insulating behavior. However, as one increases the interaction strength $U$, the conductivity switches behavior and increases with decreasing temperature: A clear evidence of interaction effects leading to de-localization of the 
down-spin species. 
 
This switching behavior of the 
conductivity with increasing interaction strength $U$ is plotted in the inset of  Fig.~\ref{quarterspindninter} where we clearly see that $\frac{d\sigma}{dT}$ for the down-spin species undergoes a sign-change at some low but finite value of $T$ thus attesting to an interaction driven metal-insulator transition.This de-localization transition is also accompanied by a net polarization. As seen in Ref.~\cite{prabudhaSSThermo}, this net polarization is purely single particle effect which can be explained via a spin-selective 
disorder induced broadening of the down-spin band as compared to the up-band. For any filling fraction away from half filling this in turn leads to a population imbalance wherein the number of down-spin particles $n_{\downarrow}$ increases, and the number of up-spin electrons $n_{\uparrow}$ decreases, thus leading to a net polarization. Further increasing the 
interaction $U$ leads to only a very weak upward renormalization of the polarization. Once again this is consistent with the results of Ref.~\cite{prabudhaSSThermo} and confirms essentially the conjecture that the net polarization can be explained via the single particle picture. 

\begin{figure}[!htbp]
\centering
\includegraphics[width=0.45\textwidth,angle=0]{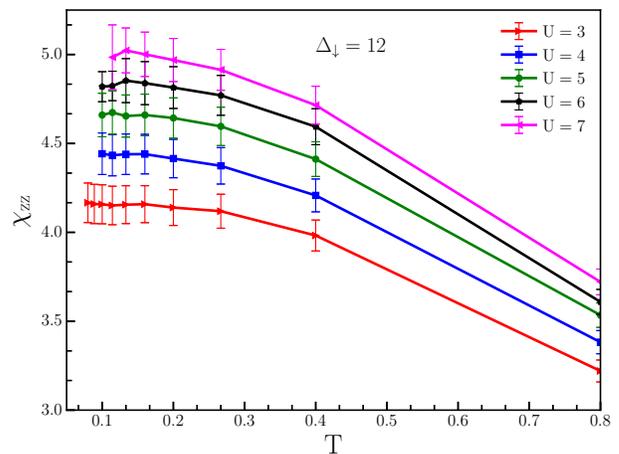}
\caption{The z-component of the ferromagnetic susceptibility versus temperature (T) plotted for various interaction strength. The plot clearly shows an enhancement of the ferromagnetic susceptibility with increasing interaction strength.}
\label{ferrosusc}
\end{figure}

Furthermore, as seen in Fig.~\ref{ferrosusc}, the delocalization transition as a function of increasing $\rm{U}$ is accompanied by a concomitant increase in the susceptibility. 
The development of an effective polarization and the increase in susceptibility is very reminiscent to the situation encountered in the RG flow equations stemming from Finkelstein's sigma model \cite{fnklstein} for interacting conventionally disordered system. In the case of the Finkelstein flow equations, in $d=2$, the delocalizing behavior is driven by an increase of the spin-triplet interaction $\gamma_t$, (or equivalently the magnetic susceptibilty) which is taken to be a signal of the onset of a ferromagnetic instability.  However, the fact that $\gamma_t \rightarrow \infty$ at a finite value of the RG scaling parameter implies a breakdown of the perturbative RG. In contrast, as elucidated above, for the case spin-selective disorder, the spin-susceptibility is just enhanced with increasing interaction and does not diverge. This potentially alludes to the fact that our spin-selective 
disordered model could be explained via Finkelstein like RG equation with suitable modification that would account for an enhanced but non-divergent susceptibility. 
Incidentally, experiments \cite{pruspudalov} in conventionally disordered two dimensional interacting electron systems also seem to suggest an enhanced but non-diverging susceptibility. Finally, we emphasize that even though our results elucidated above are for the quarter-filled lattice, there is nothing special about this filling fraction. For instance, 
for a filling fraction of $0.4$, we get similar results. This indicates that the delocalization transition highlighted above is the generic behavior whenever the system is away from the half-filled case. 

\paragraph{Conclusion:}

In our computations, it is clearly seen that sufficiently away from half-filling, even if a species of particles is localized (in our case the spin-down particles) in the bare non-interacting Hamiltonian according to Anderson localization, it is possible for the particles to delocalize (i.e., overcome the effects of Anderson localization) by interacting with another species of particles (in this case the spin-up electrons) which do not see the disorder at the bare Hamiltonian. However, clearly the lowering of the dc-conductivity of the spin-up particles with increasing interaction strength points to a scenario where the entire system does not quite delocalize but rather the original clean species start localizing (even though we do not see explicit insulating behaviour of the spin-up particles and investigation of higher interaction strengths is prohibited by the fermion sign problem) while the originally localized species start delocalizing due to the interaction. 

Interestingly, a similar study was carried out in the context of Many-Body-localization (MBL) through exact diagonalization in one dimension \cite{PRBMBL}. However, there the conclusion was each of the two species act as baths to the other and each ends up getting de-localized as a result. Here we show that this is certainly not the case in an open system in two dimensions through an exact numerical study of two dimensional interacting disordered system. On the contrary, we see an ``exchange'' of behaviour of the two species. The species that observes the bare disorder de-localizes through the interaction while the species that is a perfect band metal in the absence of interaction has reduced conductivity due to the presence of the other.

The data presented here was generated at the PG Senapathy computing facility at Indian Institute of Technology, Madras and computing facility at Indian Statistical Institute, Chennai.
Discussions with D. Belitz, R. Bhatt, K. Byczuk, J. Chalker, F. Evers, I. Gornyi, A. Mirlin, T. Vojta, D. Vollhardt and Xin Wan are gratefully acknowledged. RN thanks the Visitors program at APCTP where part of this work was accomplished.

\setcounter{equation}{0}
\setcounter{figure}{0}
\setcounter{table}{0}
\setcounter{page}{1}
\makeatletter
\renewcommand{\theequation}{S\arabic{equation}}
\renewcommand{\thefigure}{S\arabic{figure}}
\renewcommand{\bibnumfmt}[1]{[S#1]}
\renewcommand{\citenumfont}[1]{S#1}
\newpage
\makeatother
\clearpage 
\onecolumngrid

{\LARGE\bf \noindent 
Supplemental material for:\\[0.5ex] {Spin-selective metal insulator transition in two dimensions}
}
\bigskip

\noindent 
Shashikant Singh Kunwar$^1$, Prabuddha B. Chakraborty$^2$, and Rajesh Narayanan$^1$
\bigskip

{\small \noindent
$^1$ Department of Physics, Indian Institute of Technology Madras, Chennai 600036, India.\\
$^2$ Indian Statistical Institute, Chennai Centre, 110 N. Manickam Road, Rajaram Mehta Nagar, Aminjikarai, Chennai 600029, India.\\
}

\section*{S1. Methodology and Details of the Monte-Carlo Procedure}

 We utilize a DQMC scheme \cite{QMC1981}, to numerically solve the Hamiltonian of the Hubbard model in the 
 presence of spin-selective disorder. The  
 DQMC formalism initially entails the slicing of the inverse temperature $\beta$ into $L_{\tau}$ slices of 
width $\Delta\tau$ each \cite{Suzuki01111976}. The interacting electron problem is  
recast into an effective time-dependent non-interacting problem by utilizing a Hubbard-Stratonovich \cite{Hirsch1983} transformation 
in terms of an Ising-type auxillary field. The fermionic degrees of freedom are then formally integrated out. Various 
physical quantities like the Greens function, the electronic density, and the two particle correlator are calculated 
via Monte-Carlo sampling of the Ising Hubbard-Stratanovich field. The parameter space of our simulation is spanned by 
the interaction strength $U$, the disorder strength $\Delta_\downarrow$ and temperature $T$, which are all expressed in terms of the 
hopping integral $t$. The simulations are further performed at various filling fractions. The simulations are performed on a $2d$ square lattice of linear dimension $L = 10$. Thus the number of sites is $N=10^{2}$.
To incorporate the disorder averaging we have performed simulations for $80$ different disorder realizations. All the observables are obtained by performing averages over these $80$ realizations.

\section*{S2. Some Details on the dc Conductivity}

Here, we discuss some details of the approximations involved in calculating the dc conductivity: 
The dc conductivity can be calculated by taking the long wavelength($q=0$)  and $ \omega \rightarrow 0$ limit of the dynamical current-current correlation function \cite{mitcriterion}. 
\begin{equation}
\sigma_{dc} = \lim_{\omega\rightarrow 0} \frac{\mathbb{I}m\Lambda(q=0,\omega)}{\omega}
\label{eq:1}
\end{equation}

The current-current correlation function in imaginary time is connected to the dynamical current-current correlation function via the integral transform given by Eq.~(\ref{eq:2}),
\begin{equation}
\Lambda(q,\tau) = \int_{-\infty}^{+\infty} \frac{d\omega}{\pi} \frac{e^{-\omega \tau}}{1-e^{-\beta \omega}}\mathbb{I}m\Lambda(q,\omega)
\label{eq:2}
\end{equation}
The correlation function in DQMC are measured in imaginary time $\tau$. Calculating the dynamical correlation function given the correlator in imaginary time via the inversion of Eq.~(\ref{eq:2}) is not a well defined problem. 
The kernel, $k = \frac{e^{-\omega \tau}}{1-e^{-\beta \omega}} $ decays exponentially as a function of frequency. At  $\tau=\frac{\beta}{2}$ the kernel $k$ reduces to  $ \frac{1}{2 \sinh(\frac{\beta \omega}{2})}$  and it gets most of the contribution from the small $\omega$ limit. Now, in the limit $\tau=\beta/2$, $\mathbb{I}m\Lambda(0,\omega)$ can be replaced by $\omega \sigma_{dc}$ and the kernel $k$ is replaced by its value at $\tau=\frac{\beta}{2}$. \\
Once these approximations are made eq(2) becomes,
\begin{equation}
\Lambda\left(q=0,\tau=\frac{\beta}{2}\right) = \int_{- \infty}^{+ \infty} \frac{d\omega}{2 \pi} \frac{\omega \sigma_{dc}}{ \sinh(\frac{\beta \omega}{2})}
\end{equation}
After carrying out the integration,
\begin{equation}
\sigma_{dc} = \frac{\beta^{2}}{\pi} \Lambda \left( q=0,\tau=\frac{\beta}{2} \right).
\end{equation}
Now $\sigma_{dc}$ has been expressed in terms of a current-current correlation function calculated in imaginary time. Current-Current correlation function in imaginary time is calculated using DQMC and is given by, 
\begin{equation}
\Lambda(q,\tau) = \langle j_{x}(q,\tau)j_{x}(-q,\tau) \rangle.
\end{equation}
where $j_{x}(q,\tau)$ is the fourier transform of current density operator given in real space by $j_{x}(r,\tau)$, 
\begin{equation}
j_{x}(r,\tau)=e^{\frac{ \tau H}{h}}j_{x}(r)e^{-\frac{\tau H}{h}}
\end{equation}
\begin{equation}
j_{x}(r)=\frac{i e a t}{h}\sum_{\sigma} (c^{\dagger}_{r+e_{x}, \sigma}  c_{r,\sigma} - c^{\dagger}_{r,\sigma}  c_{r+e_{x}, \sigma}).
\end{equation}

\end{document}